\documentclass[12pt]{article}

\catcode`\@=11

\def\section{\@startsection{section}{1}{\z@}{3.5ex plus 1ex minus
 .2ex}{2.3ex plus .2ex}{\bf}}

\def\thesubsection{\arabic{section}.\arabic{subsection}}
\renewcommand{\subsection}[1]{\addtocounter{subsection}{1}
\vspace{2.5mm}\par\noindent {\it \thesubsection . #1}\par
 \vspace{0.5mm} }
\catcode`\@=12
\newfont{\mbm}{msbm10 scaled\magstep1}

\def\drawbox#1#2{\hrule height#2pt\hbox{\vrule width#2pt height#1pt 
 \kern#1pt\vrule width#2pt}\hrule height#2pt}

\def\Asym#1#2{\vcenter{\vbox{\drawbox{#1}{#2}\kern-#2pt\drawbox{#1}{#2}}}}

\def\be{\begin{equation}}
\def\ee{\end{equation}}
\def\ba{\begin{eqnarray}}
\def\ea{\end{eqnarray}}
\begin{document}
\begin{titlepage}
\rightline{{LPTENS 00/27}}
\rightline{{CPTH-S074.0600}}
\rightline{{LPT-ORSAY 00/61}}
\rightline{{ROM2F-2000/23}}
\rightline{{CERN-TH/2000-183}}
\rightline{{hep-th/0007090}}
\vskip 2cm
\centerline{{\large\bf Type-I strings on magnetised orbifolds and 
brane transmutation}}
\vskip 1cm
\centerline{C. Angelantonj${}^{a,b}$,
I. Antoniadis${}^{c,}$\footnote{On leave of absence from CPHT,
\'Ecole Polytechnique.}, E. Dudas${}^d$ and 
A. Sagnotti${}^{a,e}$}
\vskip 0.5cm
\centerline{\it ${}^a$ Laboratoire de Physique Th\'eorique de
l'\'Ecole Normale Sup\'erieure\footnote{Unit{\'e} mixte du CNRS et de
l'ENS, UMR 8549.}}
\centerline{\it 24, rue Lhomond, F-75231 Paris CEDEX 05}
\centerline{\it ${}^b$ Centre de Physique Th\'eorique de l'\'Ecole
Polytechnique\footnote{Unit{\'e} mixte du CNRS et de l'EP, UMR 7644.},
F-91128 Palaiseau}
\centerline{\it ${}^c$ TH-Division, CERN, CH-1211 Geneva 23}
\centerline{\it ${}^d$ LPT \footnote{Unit{\'e} mixte du CNRS, UMR
8627.}, B{\^a}t. 210, Univ. Paris-Sud, F-91405
Orsay}
\centerline{\it ${}^e$ Dipartimento di Fisica, Universit\'a di Roma
``Tor Vergata'' e INFN, Sezione di Roma 2}
\centerline{\it Via della Ricerca Scientifica 1, I-00133 Rome}
\vskip  1.0cm
\begin{abstract}
In the presence of internal magnetic fields, a D9 brane can acquire
a D5 (or anti-${\rm D}5$) R-R charge, and can therefore
contribute to the corresponding tadpole.
In the resulting vacua, supersymmetry is generically broken and 
tachyonic instabilities are present. However,
suitable choices for the magnetic fields, corresponding
to self-dual configurations in the internal space, can yield new
chiral supersymmetric vacua with gauge groups of reduced rank,
where the magnetic energy saturates, partly or fully, the negative
tension of the ${\rm O}5_{+}$ planes. These models contain Green-Schwarz
couplings to untwisted R-R forms not present in conventional
orientifolds.
\end{abstract}
\end{titlepage}

Magnetised tori were considered long ago by Witten \cite{wito32}
in the first attempt to recover 
four-dimensional chiral spectra from the low-energy field theory of 
superstrings. More recently, Bachas \cite{costas} analysed the effect of
Fradkin-Tseytlin deformations \cite{ft,acny} on open strings,
and showed how their universal magnetic couplings \cite{fpt} can
lead to chiral spectra with broken supersymmetry. However,
these models have in general
Nielsen-Olesen instabilities \cite{no}, that reflect themselves in
the emergence of tachyonic modes. This complicates the analysis, and
brings about some surprises. 
For instance, in some cases with extended (${\cal N}=2,4$)
supersymmetry, where one can analyse the potentials of the tachyonic modes, 
at the resulting minima supersymmetry is actually restored \cite{agnt}.
The constructions in \cite{wito32,costas} were both 
based on the assumption, natural at the time, 
of a vanishing instanton density for the internal magnetic field. 
However, we are now accustomed to more
general settings, that have naturally emerged from type-I vacua \cite{open}, 
where a non-vanishing instanton density is
compensated by the presence of additional branes \cite{wsi}. 
This letter is thus devoted to elucidate some peculiar effects of magnetic
deformations with non-vanishing instanton number on toroidal and
orbifold compactifications of type-I strings. As we shall see, these
can result in new vacua with unbroken supersymmetry and Chan-Paton
groups of reduced rank, where magnetised D9 branes effectively
mimic BPS D5 (anti)branes. 

It is by now well appreciated that, in non-trivial gravitational and 
gauge backgrounds,
the Wess-Zumino coupling of \cite{ghm} endows D branes with R-R
charges for forms of different degrees. It is perhaps less appreciated, 
however, that the Born-Infeld action can turn
the non-vanishing vacuum energy of suitable internal magnetic fields
into a positive tension capable of recovering the BPS bound for the
additional charges. An indirect manifestation of this 
phenomenon was recently met in \cite{abg}, where the open descendants
of some asymmetric orbifolds with 
``brane supersymmetry breaking'' \cite{bsb,bsb2,bsb3,bsb4,bsb5} were built 
using a magnetised
internal space, and where a suitable choice of internal fields played an
essential role in saturating all R-R tadpoles with only D9 branes.

Let us begin with
some intuitive field theory arguments, well captured by
the low-energy effective action for D9 branes in an internal abelian
background \footnote{The (dimensionless) magnetic fields 
used in this letter differ from the conventional 
ones by a $2 \pi \alpha'$ rescaling.},
\be
{\cal S}_9 =  - T_{(9)} \int_{{\cal M}_{10}} \!\!\!\!\! {\rm d}^{10} x \; 
e^{-\phi} \sum_{a=1}^{32} \; \sqrt{-
\det \left( g_{10} + q_a F \right)} - \mu_{(9)}
\sum_{p,a} \;
\int_{{\cal M}_{10}} \!\!\! e^{q_a F} \wedge C_{p+1} + \ldots \ \ ,
\label{s9}
\ee
where $a$ labels the types of Chan-Paton charges that couple
to the magnetic fields with strength $q_a$, 
\be
T_{(p)} = \sqrt{\pi \over 2\kappa^2 } \; \left(
2 \pi  \sqrt{\alpha'} \right)
^{3-p} = | \mu_{(p)} | \ ,
\ee
with $T$ and $\mu$ 
the tension and the R-R charge for a type-I D$p$ brane \cite{pol}, and where
$\kappa$ defines the ten-dimensional Newton constant 
$G_N^{(10)}=\kappa^2/8 \pi$.
To illustrate the phenomenon, anticipating the string construction, 
it suffices to consider the
geometry ${\cal M}_{10} = {\cal M}_6 \times T^2 \times T^2$ with
constant abelian magnetic fields $H_1$ and $H_2$ 
lying in the two internal tori.
These are effectively monopole fields, and thus satisfy the Dirac
quantisation conditions
\be
q \, H_i \, v_i = k_i \ \qquad (i=1,2) \ , \label{dirac}
\ee
where, aside from powers of
$2 \pi$, $v_i=R_i^{(1)} R_i^{(2)}/\alpha'$ are the dimensionless 
volumes of the two tori of radii $R_i^{(1)}$
and $R_i^{(2)}$, $k_i$ are the degeneracies of the 
corresponding Landau levels and $q$ is the elementary electric charge
for the system. As
anticipated, we forego the restriction in \cite{wito32,costas} 
and actually pick a pair of abelian fields aligned with
the same U(1) subgroup, so that
\ba
{\cal S}_9 &=& - \; T_{(9)} \int_{{\cal M}_{10}}\!\!\!\!\!  {\rm d}^{10} x \; 
e^{-\phi} \; \sqrt{-g_6} \; \sum_{a=1}^{32} \; 
\sqrt{(1 +  q_a^2 H_1^2 ) 
(1 + q_a^2 H_2^2 )} \nonumber \\
& & - \; 32 \, \mu_{(9)} 
\int_{{\cal M}_{10}} C_{10} \; - \; \left(2 \pi \sqrt{\alpha'}\right)^4 
\; \mu_{(9)}
\; v_1 v_2 \; H_1 \; H_2\; 
\sum_{a=1}^{32} \; q_a^2 \; \int_{{\cal M}_{6}}
C_6   \ , \label{s9special}
\ea
where ${g_6}$ denotes the six-dimensional 
space-time metric, and for simplicity
we have chosen an identity metric in the internal space. 
In particular, if the two internal fields have identical magnitudes, 
for the resulting (anti)self-dual configuration the action becomes
\ba
{\cal S}_9 &=&  - \; 32 \int_{{\cal M}_{10}} \!\!\! 
\left( {\rm d}^{10} x  \; \sqrt{-g_6}\; T_{(9)}\;
e^{-\phi} + \mu_{(9)} \; C_{10} \right) \nonumber \\
& &- \sum_{a=1}^{32} \left(\frac{q_{a}}{q}\right)^2 \; 
\int_{{\cal M}_{6}} \left( {\rm d}^{6} x \;  \sqrt{-g_6}\; 
|k_1 k_2| \; T_{(5)} \; e^{-\phi} + k_1 k_2 \; \mu_{(5)} \; 
C_{6} \right) \ . \label{s9fin}
\ea
Notice that the Dirac quantisation conditions (\ref{dirac}) 
have compensated the integration over the internal tori, while in
the second line
the additional powers of $\alpha'$ have nicely converted
$T_{(9)}$ and $\mu_{(9)}$ into $T_{(5)}$ and $\mu_{(5)}$.
Thus, a D9 brane on a magnetised $T^2 \times T^2$ indeed
mimics a D5 brane or a D5 antibrane according to whether the
orientations of $H_1$ and $H_2$, reflected by the relative sign
of $k_1$ and $k_2$, are identical or opposite.

We can now turn to the open-string description of this phenomenon.
In order to obtain a supersymmetric configuration, we should start
from an orbifold that normally requires the introduction of D5
branes. The simplest such instance is the six-dimensional
compactification on
$(T^2 \times T^2)/Z_2$ with Klein-bottle projection
\begin{equation}
{\cal K} = {\textstyle{1\over 4}} \left\{ (Q_o + Q_v) (0;0) \left[ P_1 P_2 +
W_1 W_2 \right] + 16\times 2 (Q_s + Q_c ) (0;0) \left( {\eta \over
\vartheta_4 (0)} \right)^2 \right\} \ ,
\end{equation}
that corresponds to the introduction of ${\rm O}9_+$ and ${\rm O}5_+$ planes,
and thus to a projected ${\cal N}= (1,0)$ supersymmetric closed spectrum 
with one tensor multiplet and 20 hypermultiplets.

In writing this expression, 
we have endowed the six-dimensional characters of \cite{open} with
a pair of arguments, anticipating the effect of the magnetic
deformations in the two internal tori. In general
\begin{eqnarray}
Q_o (\eta ; \zeta) \!\! &=& \!\! V_4 (0) \left[ O_2 (\eta ) O_2 (\zeta ) +
V_2 (\eta ) V_2 (\zeta ) \right] 
- C_4 (0) \left[ S_2 (\eta ) C_2 (\zeta ) +
C_2 (\eta ) S_2 (\zeta ) \right] \ ,
\nonumber
\\
Q_v (\eta ; \zeta) \!\! &=& \!\! O_4 (0) \left[ V_2 (\eta ) O_2 (\zeta ) +
O_2 (\eta ) V_2 (\zeta ) \right] 
- S_4 (0) \left[ S_2 (\eta ) 
S_2 (\zeta ) + C_2 (\eta ) C_2 (\zeta ) \right] \ ,
\nonumber
\\
Q_s (\eta ; \zeta ) \!\! &=& \!\! O_4 (0) \left[ S_2 (\eta ) C_2 (\zeta ) +
C_2 (\eta ) S_2 (\zeta ) \right] 
- S_4 (0) \left[  O_2 (\eta ) O_2 (\zeta ) +
V_2 (\eta ) V_2 (\zeta ) \right] \ ,
\nonumber
\\
Q_c (\eta ; \zeta) 
\!\! &=& \!\! V_4 (0) \left[ S_2 (\eta ) S_2 (\zeta ) +
C_2 (\eta ) C_2 (\zeta ) \right] - C_4 (0) 
\left[  V_2 (\eta ) O_2 (\zeta ) +
O_2 (\eta ) V_2 (\zeta ) \right] \ ,
\end{eqnarray}
where the four level-one O($2n$) characters are related to the
four Jacobi theta functions according to
\ba
O_{2n}(\zeta) \!\!&=&\!\! \frac{1}{2 \eta^n (\tau)} \left(
\vartheta_3^n(\zeta|\tau)
+ \vartheta_4^n(\zeta|\tau)\right) \ , \
S_{2n}(\zeta) = \frac{1}{2 \eta^n (\tau)} \left(
\vartheta_2^n(\zeta|\tau) + i^{-n} \vartheta_1^n(\zeta|\tau)\right) \
, 
\nonumber \\
V_{2n}(\zeta) \!\!&=&\!\! \frac{1}{2 \eta^n (\tau)} \left(
\vartheta_3^n(\zeta|\tau)
- \vartheta_4^n(\zeta|\tau) \right) \ , 
\
C_{2n}(\zeta) = \frac{1}{2 \eta^n (\tau)} \left(
\vartheta_2^n(\zeta|\tau)
- i^{-n} \vartheta_1^n(\zeta|\tau) \right) \ .
\ea

Whereas in \cite{wito32,costas} the
internal magnetic two-forms were chosen to satisfy
\be
{\rm tr} \; H_i \wedge H_j = 0 \ ,
\ee
here we allow for a non-vanishing instanton density, that in String Theory
is naturally compensated by additional unpaired defects (an excess of
D5 (anti)branes and/or O5 planes). In particular, as in our field
theory considerations, we take the two internal
fields aligned with the same U(1) subgroup of SO(32), a choice
that in this $Z_2$ orbifold can preserve at most a 
${\rm U}(m) \times {\rm U}(n)$ gauge group, with
$m + n = 16$. In the following, we actually restrict our attention to this
maximal case, from which other examples can be obtained via
Wilson lines or brane displacements. 

In writing the direct-channel annulus amplitude, let us 
begin by recalling \cite{acny} that a uniform
magnetic field with components $H_1$ and $H_2$ in the two internal tori 
alters the boundary conditions for open strings,
shifting their mode frequencies by
\be
z_i^{\rm L,R} \; = \; \frac{1}{\pi} \; \left[ \; \tan^{-1}( q_{{\rm L}}\,
H_i) \; + \; 
\tan^{-1}( q_{{\rm R}}\, H_i) \; \right] \ ,
\ee
where $q_{{\rm L}}$ ($q_{{\rm R}}$) denote the charges of the left (right)
end of the open string with respect to the U(1) fields $H_i$.  
A further novelty \cite{acny} 
is displayed by
``dipole'' strings, with opposite end charges, whose
oscillator modes are unaffected, but whose world-sheet coordinates undergo
a complex ``boost'', so that their Kaluza-Klein momenta $m_i$ are 
rescaled according to
\be
m_i \ \to \ \frac{m_i}{\sqrt{1 \; + \; q_a^2 H_i^2}}  \quad . \label{boost}
\ee
This rescaling ensures the consistency of the transverse-channel 
amplitudes, whose lowest-level contributions, aside from a subtlety that we
shall discuss later, are to group as usual into perfect squares.

The techniques of \cite{open} determine the direct-channel 
annulus amplitude
\begin{eqnarray}
{\cal A} &=& {\textstyle{1\over 4}} \Biggl\{ (Q_o + Q_v)(0;0) \left[
(m+\bar m)^2 P_1 P_2 + (d+\bar d)^2 W_1 W_2 
+ 2 n \bar{n} \tilde P_1 \tilde P_2 \right]
\nonumber
\\
&-& 2 (m+\bar m) (n + \bar{n}) (Q_o + Q_v )(z_1 \tau ; z_2 \tau
) {k_1 \eta \over
\vartheta_1 (z_1 \tau)} {k_2 \eta \over \vartheta_1 (z_2 \tau)} 
\nonumber
\\
&-& ( n^2 + \bar{n}^2 ) (Q_o + Q_v ) (2 z_1 \tau ; 2 z_2 \tau ) 
{2 k_1 \eta \over
\vartheta_1 (2 z_1 \tau)} {2 k_2 \eta \over \vartheta_1 (2 z_2 \tau)} 
\nonumber 
\\
&-& \left[ (m-\bar m)^2 -2 n\bar n + (d-\bar d)^2 \right] (Q_o - Q_v ) (0;0)
\left( {2\eta \over \vartheta_2 (0)}\right)^2 
\nonumber
\\
&-& 2 (m-\bar m) (n - \bar{n}) (Q_o - Q_v ) (z_1 \tau ; z_2 \tau) 
{2\eta \over \vartheta_2
(z_1 \tau)} {2\eta \over \vartheta_2 (z_2 \tau)} 
\nonumber
\\
&-& (n^2 + \bar{n}^2) (Q_o - Q_v ) (2z_1 \tau ; 2z_2 \tau)
{2\eta \over \vartheta_2
(2z_1 \tau)} {2\eta \over \vartheta_2 (2z_2 \tau)} 
\nonumber
\\
&+& 2 (m+\bar m ) (d+\bar d) (Q_s + Q_c) (0;0) \left({\eta \over
\vartheta_4 (0)}\right)^2 
\nonumber
\\
&+& 2 (d+\bar d)(n + \bar{n})(Q_s + Q_c) (z_1 \tau ; z_2 \tau)
{\eta \over \vartheta_4
(z_1 \tau )} {\eta \over \vartheta_4 (z_2 \tau )}
\nonumber
\\
&-& 2 (m-\bar m) (d - \bar d) (Q_s - Q_c )
(0;0) \left( {\eta \over \vartheta_3 (0)}\right)^2 \label{annsusy}
\\
&-& 2 (d-\bar d) (n - \bar{n}) (Q_s - Q_c) (z_1 \tau ; z_2 \tau)
 {\eta \over \vartheta_3
(z_1 \tau )} {\eta \over \vartheta_3 (z_2 \tau )} \Biggr\} \ ,
\nonumber
\end{eqnarray}
and the corresponding M\"obius amplitude
\begin{eqnarray}
{\cal M} &=& -{\textstyle{1\over 4}} \Biggl\{ 
(\hat Q_o + \hat Q_v )(0;0) \left[ (m+\bar m) P_1 P_2 + (d+\bar d) W_1
W_2 \right]
\nonumber
\\
&-& ( n + \bar{n}) (\hat Q_o + \hat Q_v ) (2z_1 \tau ; 2z_2 \tau) {2 k_1
\hat\eta \over \hat \vartheta_1 (2z_1\tau)} {2 k_2
\hat\eta \over \hat \vartheta_1 (2z_2\tau)}
\nonumber
\\
&-& \left( m+ \bar m + d + \bar d \right) (\hat Q_o - \hat Q_v )(0;0) \left(
{2\hat\eta \over \hat \vartheta_2 (0)}\right)^2 \label{mobsusy}
\\
&-& (n + \bar{n}) (\hat Q_o - \hat Q_v ) (2 z_1 \tau ; 2 z_2 \tau )
{2\hat\eta \over \hat\vartheta_2 (2z_1\tau)}
{2\hat\eta \over \hat\vartheta_2 (2z_2\tau)} \Biggr\} \ .
\nonumber
\end{eqnarray}
Here we have actually resorted to a shorthand notation,
where the arguments $z_i$  ($2z_i$) are associated to strings with one
(two) charged ends.
Moreover, both the imaginary modulus 
$\frac{1}{2} i t$ of ${\cal A}$ and the complex modulus $
\frac{1}{2} + \frac{1}{2} i t$ of ${\cal M}$ are denoted
by the same symbol $\tau$, although
the proper ``hatted'' contributions to the M\"obius amplitude are explicitly
indicated. $P_i$ and $W_i$ are conventional momentum
and winding sums for the two-tori, while a ``tilde'' denotes a sum with
momenta ``boosted'' as in (\ref{boost}). Finally, 
$d$ (together with its conjugate $\bar{d}$) is the
Chan-Paton multiplicity for the D5 branes, while 
$m$ and $n$ (together with their conjugates $\bar{m}$ and $\bar{n}$) 
are Chan-Paton multiplicities for the D9 branes. For the sake
of brevity, several terms with
opposite U(1) charges, and thus with opposite $z_i$ arguments, 
have been grouped together, using the symmetries of 
the Jacobi theta-functions.

For generic magnetic fields, the open spectrum is indeed non-supersymmetric
and develops Nielsen-Olesen instabilities \cite{no}. As emphasised in
\cite{costas}, the emergence of these tachyonic modes can be ascribed to the
magnetic couplings of the internal components of gauge fields. 
For instance, small magnetic fields
affect the mass formula for the untwisted string modes according to
\be
\Delta M^2 \; = \; \frac{1}{2 \pi \alpha'} \; \sum_{i=1,2} 
\left[  (2 n_i + 1) |(q_{\rm L} + q_{\rm R}) H_i| + 
2 (q_{\rm L} + q_{\rm R}) \Sigma_i H_i \right] \ ,
\ee
where the first term originates from the Landau levels and
the second from the magnetic moments of the spins $\Sigma_i$. 
For the internal
components of the vectors, the magnetic moment coupling generally
overrides the zero-point contribution, leading to tachyonic modes,
unless $|H_1|=|H_2|$,
while for spin-$\frac{1}{2}$ modes it can at most compensate it. 
On the other hand, for twisted modes the zero-point contribution
is absent, since ND strings have no Landau levels. In this case
the low-lying space-time fermions, that originate from the fermionic part 
$S_4 O_4$ of $Q_s$, are scalars in the internal space and
have no magnetic moment couplings. However, their bosonic partners, 
that originate from $O_4 C_4$,
are affected by the magnetic deformations and have mass shifts
$\Delta M^2 \sim \pm (H_1 - H_2)$.  Therefore, 
if $H_1=H_2$ all tachyonic instabilities
are indeed absent. Actually, with this choice
the supersymmetry charge, that belongs to $C_4 C_4$, 
is also unaffected\footnote{Type-II
branes at angles preserving some supersymmetry 
were previously considered in \cite{bdl}.  After T-dualities, these 
can be related to special choices for the internal 
magnetic fields. Type I toroidal models, however, can not lead to 
supersymmetric configurations, since the resulting R-R tadpoles
require the introduction of antibranes.}. Therefore,
a residual supersymmetry is present for the
entire string spectrum, and indeed, 
using the Jacobi identities for non-vanishing arguments \cite{ww}, 
one can see that for $z_1=z_2$ 
both ${\cal A}$ and ${\cal M}$ vanish identically. Still, the 
resulting supersymmetric models are rather peculiar, 
as can be seen from the deformed tadpole conditions, 
to which we now turn.

Let us begin by examining the untwisted R-R tadpole conditions. 
For $C_4 S_2 C_2$ one finds
\be
\left[ m+\bar m + n + \bar{n} - 32 +  q^2  H_1 H_2 (n +
\bar n ) \right] \sqrt{v_1 v_2} 
+  {1\over \sqrt{v_1 v_2}} \left[ d+\bar d
- 32\right]  =0 \ , \label{rrutad}
\ee
aside from terms that vanish after identifying the multiplicities
of conjugate representations $(m,\bar{m})$, $(n,\bar{n})$ and $(d,\bar{d})$.
The additional (untwisted) R-R tadpole conditions 
from $Q_o$ and $Q_v$ are compatible with (\ref{rrutad}) and do not add further 
constraints. This expression reflects the familiar Wess-Zumino
coupling of eq. (\ref{s9}), and therefore the various powers of $H$
correspond to R-R forms of different degrees. 
In particular, as we anticipated in our field theory discussion,
the term bilinear in the magnetic fields has a very
neat effect: it charges the D9 brane with respect to the
six-form potential. This can be seen very clearly making use of the
quantisation condition  (\ref{dirac}), that turns the tadpole
conditions (\ref{rrutad}) into
\begin{eqnarray}
m+\bar m + n + \bar n &=& 32 \ ,
\nonumber
\\
k_1 k_2 (n + \bar n )  + d + \bar d &=& 32 \ . \label{urrt}
\end{eqnarray}
Thus, if $k_1 k_2 > 0$ the D9 branes indeed acquire the
R-R charge of $|k_1 k_2|$ D5 branes, while if $k_1 k_2 < 0$ 
they acquire the R-R charge of as many D5 antibranes, in agreement
with eq. (\ref{s9fin}).

The untwisted NS-NS tadpoles exhibit very nicely their relation to the 
Born-Infeld term in (\ref{s9}). For instance, the dilaton tadpole
\be
\left[ m+\bar m + (n + \bar n) \sqrt{\left( 1 +  q^2 H_1^2 \right) 
\left( 1 + q^2 H_2^2 \right) }  -32 \right] \sqrt{v_1 v_2} 
+ {1\over \sqrt{v_1 v_2}} \left[ d +
\bar d - 32 \right] \label{diltad}
\ee
originates from $V_4 O_2 O_2$, and can be clearly linked to
the derivative of the integrand of ${\cal S}_9$, specialised
to the form (\ref{s9special}), with respect to
$\phi$. On the other hand, the volume of the first internal torus 
originates from $O_4 V_2 O_2$, and the corresponding tadpole,
\be
\left[ m+\bar m + (n + \bar n) \, {1 - q^2 H_1^2 \over 
\sqrt{ 1 + q^2 H_1^2 }}\,
\sqrt{ 1 + q^2 H_2^2 } -32 \right] \sqrt{v_1
v_2}
- {1\over \sqrt{v_1 v_2}} \left[ d+\bar
d - 32 \right]  \ , \label{mettad}
\ee
can be linked to the derivative of the Born-Infeld action 
in (\ref{s9}) with respect to
the corresponding breathing mode. A similar result holds for the
volume of the second torus, with the proper interchange
of $H_1$ and $H_2$. For the sake of brevity, we have 
omitted in these NS-NS tadpoles all terms
that vanish using the constraint $n = \bar n $. However, the full expression 
of (\ref{mettad}) is
rather interesting, since, in contrast with the usual structure
of unoriented string amplitudes, it is {\it not} a perfect square. 
This unusual feature can be ascribed to the behaviour of the 
internal magnetic fields under time
reversal. Indeed, as stressed long ago in \cite{cardy}, these
transverse-channel amplitudes involve a time-reversal operation ${\cal T}$, 
and are thus of the form $\langle {\cal T} (B) | q^{L_0} | B \rangle$. 
Differently from the usual quantum mechanical amplitudes, this type of 
expression is generally a bilinear, rather than
a sesquilinear, form. This, however, is not true in
the present examples, where additional signs are introduced by the
magnetic fields, that are odd under time reversal. As a result, in
deriving from factorisation the M\"obius amplitudes for these models, 
it is crucial to add the two contributions $\langle {\cal T} (B) 
| q^{L_0} | C \rangle $ and $\langle {\cal T} (C) | q^{L_0} |
B \rangle $, that are different and
effectively eliminate the additional terms from the transverse-channel.

Both (\ref{mettad}) and the dilaton tadpole (\ref{diltad}) simplify
drastically in the interesting case $H_1=H_2$ where, using the
Dirac quantisation conditions (\ref{dirac}), they become
\be
\left[ m+\bar m + n + \bar n -32 \right] \sqrt{v_1 v_2} 
\mp {1\over \sqrt{v_1 v_2}} \left[ k_1 k_2 (n
+ \bar n) + d +
\bar d - 32 \right] \ .
\ee
Thus, they both vanish, as they should, in these supersymmetric 
configurations, once the corresponding R-R tadpole conditions 
(\ref{urrt}) are enforced.

The twisted R-R tadpoles
\begin{equation}
15 \left[ {\textstyle{1\over 4}} (m-\bar m + n -\bar n )
\right]^2
+ \left[ {\textstyle{1\over 4}} (m-\bar m + n - \bar n ) - (d-\bar d)
\right]^2 
\end{equation}
originate from the sector  $S_4 O_2 O_2$,
whose states are scalars in the internal space. They 
reflect very neatly the distribution of branes among the sixteen
fixed points, only one of which accommodates D5 branes in our
examples,
are not affected by the magnetic fields, and vanish identically for the given 
unitary gauge groups. In general these
breaking terms, that originate from twisted modes flowing in the 
transverse channel, can be linked to internal curvature contributions to 
the Wess-Zumino term, here localised at the fixed points: this 
is actually the reason for the presence of D9 and D5 terms
in the same expression in orbifold models. 
The corresponding NS-NS tadpoles, originating from the
$O_4 S_2 C_2$ and $O_4 C_2 S_2$ sectors, 
are somewhat more
involved, and after the identification of conjugate multiplicities
are proportional to
\be
{q \, (H_1 - H_2 )  \over \sqrt{ (1 + q^2 H_1^2 ) 
(1 + q^2 H_2^2 ) }} \quad .
\ee
They clearly display new couplings for twisted
NS-NS fields that, to the best of our knowledge, were not
previously exhibited. Notice that, as expected,  for $H_1 = H_2$
these twisted tadpoles also vanish.

We can now describe  some supersymmetric models corresponding to
the special choice $H_1=H_2$. It suffices to confine our
attention to the case $k_1 = k_2 =2$, the minimal
Landau-level degeneracies allowed on this $Z_2$ orbifold. 
Although the projected closed spectra of all the resulting models 
are identical, and comprise the ${\cal N}=(1,0)$ 
gravitational multiplet, together with one tensor multiplet and
twenty hypermultiplets, the corresponding open spectra are quite different
from the standard ones, with a maximal
gauge group of rank 32, ${\rm U}(16)|_9 \times {\rm U}(16)|_5$ 
\cite{bs,gp}.
Still, they are all free of irreducible gauge and gravitational anomalies,
consistently with the vanishing of all R-R tadpoles \cite{pc}.

A possible solution to the R-R tadpole conditions is
$m=13$, $n=3$, $d=4$, that corresponds to a gauge group of rank 20,
${\rm U}(13)|_9 \times {\rm U}(3)|_9 \times {\rm U}(4) |_5$, with
charged hypermultiplets in the representations $(78 + \overline{78},1;1)$,
in five copies of the $(1,3  + \overline{3};1)$, in one copy of the
$(1,1;6  + \overline{6})$, in four copies of the
$(\overline{13},3;1)$, in one copy of the $(13,1;\overline{4})$ and
in one copy of the  $(1,\overline{3};4)$. 
Alternatively, one can take $m=14$, $n=2$, $d=8$,
obtaining a gauge group of rank 24,
${\rm U}(14)|_9 \times {\rm U}(2)|_9 \times {\rm U}(8) |_5$. The
corresponding matter comprises charged
hypermultiplets in the $(91 + \overline{91},1;1)$, 
in one copy of the $(1,1;28  + \overline{28})$, in four copies of the
$(\overline{14},2;1)$, in one copy of the $(14,1;\overline{8})$,
in one copy of the  $(1,2;8)$, and in five copies of the 
$(1,1 + \overline{1},1)$. On the other hand, for
$m=12$, $n=4$, and thus $d=0$. This is a rather unusual supersymmetric 
$Z_2$ model {\it without} D5 branes, with a gauge group of rank 16, 
${\rm U}(12) \times {\rm U}(4)$, and charged 
hypermultiplets in the representations $(66 + \overline{66},1)$,
in five copies of the $(1,6 + \overline{6})$, and in four copies of
the $(\overline{12},4)$. 
A distinctive feature of these
spectra is that some of the matter occurs in multiple families. 
This peculiar phenomenon is a consequence of the multiplicities of
Landau levels, that in these $Z_2$ orbifolds are multiples of two for
each magnetised torus. Moreover,
it should be appreciated that, in general, the
rank reduction for the gauge group is not by powers of two as
in the presence of a quantised antisymmetric tensor \cite{tens,bsb2}.
Actually, these are not the first
concrete examples of brane transmutation in type I vacua but, to
the best of our knowledge, they
are the first supersymmetric ones.
$Z_2$ orientifolds without D5 branes have recently appeared in
\cite{abg}, where magnetised fractional D9 branes have been used to build
six-dimensional asymmetric orientifolds with ``brane
supersymmetry breaking''. 

One can also consider similar deformations of the model of \cite{bsb},
that has an ${\cal N}=(1,0)$ supersymmetric bulk spectrum with 
17 tensor multiplets
and four hypermultiplets. This alternative projection, allowed by the
constraints in \cite{pss}, introduces ${\rm O}9_{+}$
and ${\rm O}5_{-}$ planes and thus requires, for consistency, an open sector
resulting from the simultaneous presence of D9 branes and D5 antibranes,
with ``brane supersymmetry breaking''. A magnetised torus
can now mimic D5 antibranes provided $H_1 = - H_2$, and one can then build
several non-tachyonic configurations as in the previous case\footnote{
There is a subtlety here. The different GSO projections for
strings stretched between a D9 brane and a D5 antibrane would associate
the low-lying  twisted ND bosons to the characters $O_4 S_2(z_1) S_2(z_2)$
and $O_4 C_2(z_1) C_2(z_2)$, and thus now
the choice $H_1 = - H_2$ would eliminate all tachyons even in the presence
of D5 antibranes.}. A particularly
interesting one corresponds to a vacuum configuration without D5 
antibranes, where the ${\rm O}5_{-}$ charge is fully saturated by
magnetised D9 branes. The resulting annulus and M\"obius amplitudes
can be obtained deforming the corresponding ones in \cite{bsb}, and read
\begin{eqnarray}
{\cal A} &=& {\textstyle{1\over 4}} \Biggl\{ (Q_o + Q_v)(0;0) \left[
(m_1+m_2)^2 P_1 P_2 + 2 n \bar{n} \tilde P_1 \tilde P_2 \right]
\nonumber
\\
&-& 2 (m_1+ m_2) (n + \bar{n}) (Q_o + Q_v )(z_1 \tau ; z_2 \tau
) {k_1 \eta \over
\vartheta_1 (z_1 \tau)} {k_2 \eta \over \vartheta_1 (z_2 \tau)} 
\nonumber
\\
&-& ( n^2 + \bar{n}^2 ) (Q_o + Q_v ) (2 z_1 \tau ; 2 z_2 \tau ) 
{2 k_1 \eta \over
\vartheta_1 (2 z_1 \tau)} {2 k_2 \eta \over \vartheta_1 (2 z_2 \tau)} 
\nonumber 
\\
&+& \left[ (m_1-m_2)^2 + 2 n\bar n \right] (Q_o - Q_v ) (0;0)
\left( {2\eta \over \vartheta_2 (0)}\right)^2 
\nonumber
\\
&+& 2 (m_1-m_2) (n + \bar{n}) (Q_o - Q_v ) (z_1 \tau ; z_2 \tau) 
{2\eta \over \vartheta_2
(z_1 \tau)} {2\eta \over \vartheta_2 (z_2 \tau)} 
\nonumber
\\
&+& (n^2 + \bar{n}^2) (Q_o - Q_v ) (2z_1 \tau ; 2z_2 \tau)
{2\eta \over \vartheta_2
(2z_1 \tau)} {2\eta \over \vartheta_2 (2z_2 \tau)} \Biggr\} \quad ,
\label{annsusyb}
\end{eqnarray}
and
\begin{eqnarray}
{\cal M} &=& -{\textstyle{1\over 4}} \Biggl\{ 
(m_1 + m_2) ( \hat Q_o + \hat Q_v )(0;0)   P_1 P_2 
\nonumber
\\
&-& ( n + \bar n ) (\hat Q_o + \hat Q_v ) (2z_1 \tau ; 2z_2 \tau) {2 k_1
\hat\eta \over \hat \vartheta_1 (2z_1\tau)} {2 k_2
\hat\eta \over \hat \vartheta_1 (2z_2\tau)}
\nonumber
\\
&+& \left( m_1+ m_2 \right) (\hat Q_o - \hat Q_v )(0;0) \left(
{2\hat\eta \over \hat \vartheta_2 (0)}\right)^2 \label{mobsusyb}
\\
&+& (n + \bar n ) (\hat Q_o - \hat Q_v ) (2 z_1 \tau ; 2 z_2 \tau )
{2\hat\eta \over \hat\vartheta_2 (2z_1\tau)}
{2\hat\eta \over \hat\vartheta_2 (2z_2\tau)} \Biggr\} \quad .
\nonumber
\end{eqnarray}
In extracting the massless spectra of this class of models, 
it is important to notice that, at the
special point $H_1 = - H_2$, all bosons from $Q_o$
with non-vanishing arguments and all fermions from
$Q_v$ with non-vanishing arguments become massive. 
As a result, all massless fermions arising from strings affected by the
internal magnetic fields have a reversed chirality, precisely as
demanded by the cancellation of all irreducible anomalies. 
For $|k_1 |= |k_2 | = 2$, one can obtain
a gauge group ${\rm SO} (8) \times {\rm SO} (16) \times {\rm U} (4)$
and, aside from the corresponding ${\cal N}= (1,0)$ vector multiplets,
the massless spectrum contains a hypermultiplet in the representation 
$(8,16,1)$, eight scalars in the $(1,16,4+\overline{4})$, two
left-handed spinors in the $(8,1,4+\overline{4})$, and twelve scalars 
and five left-handed spinors in the $(1,1,6+\overline{6})$. 
Clearly, supersymmetry is explicitly broken on the magnetised D9 branes. 
Still, the resulting dilaton potential is effectively localised 
on the ${\rm O}5_-$ plane, since it scales with the internal volume 
as in the undeformed model of \cite{bsb}.

These configurations present another interesting novelty:
they have generalised Green-Schwarz couplings \cite{gs,ggs} involving
gauge fields and untwisted R-R forms, of the type
\be
{\cal S}_{\rm GS} \sim \sum_i \; \int \epsilon^{\mu_1 \ldots \mu_6}\;
\epsilon^{I_1 \ldots I_4} 
\; C_{I_1 I_2 \mu_3 \mu_4 \mu_5 \mu_6} \; {\rm tr} \left( 
F_{\mu_1 \mu_2} H_{I_3 I_4}^i \right)
 \ , \label{wzfst}  
\ee
while standard orientifolds do not \cite{madrid}.
In six dimensions, these four-forms are actually dual to axions $a_{IJ}$,
and therefore this coupling can be rewritten in the form
\be
{\cal S}_{\rm GS} \sim \sum_i \; \int {\rm tr} (A_\mu Q^i)\; H^i_{IJ} \;
\partial^{\mu}
a^{IJ} \ ,
\ee
where $Q^i$ denote the group generators associated to
the internal magnetic fields. 
Thus, additional U(1) gauge fields can acquire
mass by a generalisation of the mechanism in \cite{wito32,dsw}, that
in type-I strings generally involves several R-R forms. 
The (non-universal) axions
involved in these Higgs mechanisms are
the linear combinations $H^i_{IJ} \; a^{IJ}$.

A convenient way to recover these couplings uses, as in \cite{fabre}, 
a space-time magnetic background ${\cal F}$ that, when
introduced in the string amplitudes
(\ref{annsusy}) and (\ref{mobsusy}), deforms the space-time
theta-functions according to
\be
{1 \over \eta^2} \frac{\vartheta_\alpha (0|\tau)}{\eta (\tau)}
\ \rightarrow \ (q_{\rm L}+q_{\rm R}) {\cal F}  
\tau \; \frac{\vartheta_\alpha (\epsilon \tau 
|\tau)}{\vartheta_1  (\epsilon\tau |\tau )} \ , \label{1.8}
\ee  
with $\pi \epsilon = \tan ^{-1} (q_{\rm L} {\cal F} ) + 
\tan ^{-1} (q_{\rm R} {\cal F} )$.
As a result, the untwisted R-R tadpoles are modified, and become 
\be
\left[ m+\bar m + n + \bar{n} - 32 + 
q^2 (H_1 H_2 +{\cal F} H_1 + {\cal F } H_2) (n +\bar n ) \right] 
\sqrt{v_1 v_2} \pm {1\over \sqrt{v_1 v_2}} \left[ d+\bar d
- 32\right] \ .   \label{1.9}
\ee
Using the tadpole conditions (\ref{urrt}) and the Dirac quantisation
conditions (\ref{dirac}), the terms linear in the space-time
magnetic field identify the new Green-Schwarz couplings of
eq. (\ref{wzfst}), needed to dispose of the new anomalous U(1) factors.

In conclusion, we have seen how in type I vacua a
non-vanishing (anti)\-instanton density can be used to mimic BPS D5
(anti)branes, and we have exhibited some models with new distinctive
features. These include supersymmetric $T^4/Z_2$ compactifications 
{\it without} D5 branes, or with gauge 
groups of unusual rank, that display new Green-Schwarz couplings of
untwisted R-R forms.
Several examples of this type can be constructed, both in six and in
four dimensions. For instance, in the $Z_3$ orientifold of \cite{chiral}
magnetic deformations allow the introduction of a {\it net} number of 
D5 (anti)branes, a setting to be contrasted with the models 
of \cite{bsb3,bsb4,ibanez}, that only involve D5 brane-antibrane pairs.
Models with ``brane supersymmetry
breaking'', in particular with
additional brane-antibrane pairs,  develop NS-NS tadpoles. 
These tadpoles are not eliminated by the magnetic deformation, and 
typically result in 
potentials that, although of run-away type for the dilaton, 
can in some cases stabilise some geometric moduli \cite{bsb4}. 
Their presence requires a background redefinition \cite{fs}, 
that was recently constructed explicitly in \cite{dm} for the model in
\cite{sugimoto}. In general, these vacua
correspond to supergravity models frozen in phases of broken
supersymmetry, where the presence of (lower-dimensional)
non-supersymmetric couplings renders the field equations naively inconsistent,
in complete analogy with ordinary gauge theories frozen in a Higgs
phase. Although similar features were previously met in the
anomalous Green-Schwarz 
couplings of \cite{ggs}, the peculiar supergravity models 
resulting from ``brane supersymmetry breaking'' 
clearly deserve further investigation.

\vskip 24pt

\noindent
{\bf Acknowledgements.} We are grateful to P. Bain, F.S. Hassan, 
E. Kiritsis and R. Minasian for interesting discussions.
C.A. is supported by the ``Marie-Curie'' fellowship
HPMF-CT-1999-00256. This work was supported in part by the EEC 
contract HPRN-CT-2000-00122, in part by the EEC contract
ERBFMRX-CT96-0090, in part by the EEC contract ERBFMRX-CT96-0045 and
in part by the INTAS project 991590.

\end{document}